\documentclass[12pt]{article}
\usepackage{graphicx}
\begin{document}
{\bf Anomalous diffusion at percolation threshold in high dimensions on
$10^{18}$ sites}

\bigskip
Dirk Osterkamp, Dietrich Stauffer and Amnon Aharony$^{1}$

\bigskip
Institute for Theoretical Physics, Cologne University, 
D-50923 K\"oln,  Euroland

\bigskip
\noindent
$^1$ visiting from School of Physics and Astronomy, Raymond and Beverly Sackler 
Faculty of Exact Sciences, Tel Aviv University, Ramat Aviv, Tel Aviv 69978, 
Israel
\medskip

\noindent
e-mail: osterkamp@gmx.de, stauffer@thp.uni-koeln.de, aharony@post.tau.ac.il

\bigskip
{\small Using an inverse of the standard linear congruential random number
generator, large randomly occupied lattices can be visited by a random walker
without
having to determine the occupation status of every lattice site in advance.
In seven dimensions, at the percolation threshold with $L^7$ sites and $L \le 
420$, we confirm the expected time-dependence of the end-to-end distance
(including the corrections to the asymptotic behavior).  }
\bigskip

\bigskip

For transport in disordered media, like the ant-in-the-labyrith model of random
percolation \cite{benav,gefen,havlin}, one usually first constructs the 
disordered medium, and then starts the transport medium. In a hypercubic lattice
of linear size $L$ in $d$ dimensions, one first has to decide about the status
% "fill" replaced by "decide about the status of"
of $L^d$ sites. In two
dimensions this is efficient since for long enough time every site is visited.
In higher dimensions, however, only a small fraction of the lattice is visited
and it is more efficient to determine the status of a site only when it is 
visited first, keeping this status fixed for later visits. We achieve this aim
here for the case that every site is randomly allowed for the random walker
with probability $p$ and forbidden with probability $1-p$. We apply this method
to seven-dimensional percolation \cite{ziff,grass} at its threshold $p_c = 0.088951$
where an infinite network of allowed sites is just possible \cite{books}. We
found good agreement of the new method $(L \le 420)$ with the traditional one
$(L = 23)$, while for smaller $L$ finite-size effects are seen.

Lattice sites $i = 1,2, \dots L^d$ are numbered in helical order such that the
neighbours of site $i$ are $i \pm 1, \; i \pm L, \; i \pm L^2, \dots, i \pm 
L^{d-1}$. In the normal technique, we fill one site after the other using 
consecutive random numbers $0 < x_i < 1$; if $x_i < p$ the site is allowed. 
With a simple linear congruential random number generator like
$$ j_{i+1} = k j_i \quad {\rm mod}\; m; \quad m = 2^n$$
for integers $j_i$ instead of real numbers $x_i$,
% confusing notation x for both real and integer avoided: real x, integer j
on computers with at least $n$ bits per word, this method means that a step
of the random walker to the right ($i \rightarrow i+1$) corresponds to a 
multiplication of $j_i$ 
by $k$, a step to the left to a multiplication \cite{jones} by the inverse 
$k$* such that $k k$* = 1 modulo $m$. This inverse exists if and only if $k$ and
$m$ are relatively prime; then $k$* can easily be calculated by the extended
Euclid algorithm \cite{knuth}. Tables of pairs $k, k$* were given by
L'Ecuyer \cite{ecu} or can be easily calculated by a Fortran program available
from DO; for example $k$ = 16807 works with $k$* = 1278498327 for $n=32$ and
$k$* = 4409460005719528983 for $n = 63$. We used on a 64-bit Cray-T3E

$k = 3512401965023503517, \quad k$* = 3753721746144068021,

\noindent
table 5 of \cite{ecu}. If a step upwards, $i \rightarrow
i+L$, corresponds to $L$ multiplications with $k$ (modulo $m$), then a 
downwards step  corresponds to $L$ multiplications with $k$* (modulo $m$), 
i.e. $L$ pseudo-divisions by
$k$. The other directions involve higher powers of $L$. In this way an arbitrary
walk can be followed in a reproducible way by the proper number of 
multiplications with $k$ and $k$*, always using the modulo restriction.
The number of sites is then only limited by the period of the random number
generator; we used a multiplier \cite{ecu} with good spectral results and 
maximal period $2^{61}$. This method can be extended to the more general 
generator $j_{i+1} = k j_i + c \; {\rm mod} \, m$. 

We use helical boundary conditions in $d-1$ directions and unlimited extent in 
the remaining direction, i.e. site $i$ is thought to have as its right neighbour
the site $i+1$ even if $i$ is an integer multiple of $L$ and lies at the right
boundary. And if due to a large jump the new
site has $i \le 0$ or $i > L^d$ we do not have to put $i$ back into the
traditional interval $1 \le i \le L^d$ through $i \rightarrow i \pm L^d$ since
the new method does not actually store an array with index $i$ from 1 to
$L^d$. In this sense our lattice size is infinite in one direction. 
(It is easiest to imagine a planar lattice, with sites numbered in a typewriter
fashion.) 

In our 32-bit Fortran program, instead of the modulo function 
we used the automatic omission of leading bits if a
product of two integers gives an integer with more than 32 bits. Since
Fortran does not have the unsigned long integer type of C, the resulting 
products between $-2^{31}$ and $+2^{31}$ are often negative. We tested by
comparison with the traditional method that the new technique works 
nevertheless. For 64 bits, we masked off the leading bit, thus
working modulo $2^{63}$ with only positive integers. This program including all
{\tt shmem} commands for communication between different Cray processors 
still fits onto 68 lines, Fig. 1. It can also be used for less than seven
dimension by reducing the parameter {\tt idim}; we made test runs with 
$1,100,000,000^2, \; 1,050,000^3, \; 32800^4, \; 4100^5, \; 1001^6$ sites,
as well as $180^8$.

\begin{verbatim}
c     diffusion in d < 9 dimensions on random lattice; < 10^18 to avoid repeats
      implicit none
      real p,factor,av(30),avs(30),time
      integer L,idim,maxtime,num,i,iwalker,maxt,iter,itime
      integer idir,isum,number,node
      integer*8 ishift(0:15),mult(0:15),ip,imod,iseed,isee2,ibm,ibmj,new
      integer shmem_n_pes, shmem_my_pe, info, iadd, barrier
      common /t3e/ av, avs

      data L/420/,idim/7/,p/0.088951/,maxtime/17/,num/1000/
      data av/30*0.0/,iseed/50923/,isee2/4711/
      data imod/'7FFFFFFFFFFFFFFF'x/
      node=shmem_my_pe()
      number=shmem_n_pes()
      time=irtc()*3.33e-9
      if(node.eq.0) print *,p,L,idim,iseed,maxtime,num,number
 
      ibm =2*iseed-1
      ibmj=2*isee2-1
      do 1 iter=0,node
        ibm =ibm *65539
 1      ibmj=ibmj*65539
      mult(0)=3512401965023503517
      mult(1)=3753721746144068021
      do 2 i=2,idim*2-1
 2      mult(i)=iand(mult(i-2)**L,imod)
      ip=p*imod
      factor=1.0/(number*num)

      do 8 iwalker=1,num
        do 3 i=1,idim
          ishift(2*i-2)=0
 3        ishift(2*i-1)=0
 4      ibmj=iand(ibmj*16807,imod)
        if(ibmj.gt.ip) goto 4
        maxt=2
        do 8 iter=1,maxtime
            if(iter.gt.2) maxt=maxt*2
c           evaluation of distances only if time is power of two
            do 6 itime=1,maxt
 5            ibm=ibm*16807
              idir=ishft(ibm,-60)
              if(idir.ge.2*idim) goto 5
              new=iand(ibmj*mult(idir),imod)
              if(new.gt.ip) goto 6
              ibmj=new
              ishift(idir)=ishift(idir)+1
 6          continue
            isum=0
            do 7 i=1,idim
 7            isum=isum+(ishift(2*i-2)-ishift(2*i-1))**2
            av(iter)=av(iter)+isum
 8    continue

      info = barrier()
      if(node.gt.0) stop
      if (number.gt.1) then
        do 9 iadd = 1,number-1
          call shmem_get(avs,av,maxtime,iadd)
          do 9 i=1,maxtime
 9          av(i)=av(i)+avs(i)
      endif
      do 10 iter=1,maxtime
 10     print *, iter,av(iter)*factor,2**iter
      time = irtc() * 3.33e-9 -time
      print *,'CPU= ',4*time,number
      stop
      end 
\end{verbatim}
 
\begin{figure}[hbt]
\begin{center}
\end{center}
\caption{ Fortran program for Cray-T3E parallel computing, averaging over 
{\tt number} processors each simulating {\tt num} random walkers. 
Without the complications of parallel computation, loops 1 and 9 as well as 
all commands involving {\tt shmem, barrier, node, number, avs} can be omitted.}
\end{figure}

\bigskip
Now we repeat the standard scaling theory \cite{benav,gefen,havlin,books} for
the average end-to-end distance $r(t,p)$ and enlarge it by predicting 
correction terms.

Let $R_s, \; <r_s(t)>, \; <r(t)>\; $ be the average radius of gyration of a 
cluster
containing $s$ sites, the average end-to-end distance travelled on such a
cluster in time $t$, and this distance averaged over all starting points on
all clusters. Analogous notations hold for the squared distances; these
averages $<r^2(t)>$ over the squares do in general not scale like the squares of
 the
averages $<r(t)>$. The number
of clusters containing $s$ sites each is $n_s =
s^{-\tau}f[(p_c-p)s^\sigma]$ with the standard critical exponents $\sigma =
1/(\beta \delta), \; \tau = 2 + 1/\delta$. In general dimensions $d$, we have
$R_s = s^{\sigma \nu} h[(p_c-p)s^\sigma]$ with the correlation length
exponent $\nu$, and assume
$$ <r_s^2(t)> \; = R_s^2 \; g(t/s^x, R_s/s^\sigma) \quad .$$
Right at $p=p_c$ this assumption simplifies to
$$ <r_s^2(t)> \; = R_s^2 \; G(t/s^x) \quad ,$$ 
(varying as $R_s^2$ for $t \gg s^x$, when every site on the cluster has been 
visited many times, and as $t^{2\sigma\nu/x} = t^{2/d_w}$, similar to the 
anomalous behavior on the infinite cluster, for $1 \ll t \ll s^x$). 
For $p < p_c$ and for sufficiently long times
one has only the former behavior: the random walker has visited the whole 
finite cluster, and thus $r_s(t=\infty) \simeq
R_s$. Here, $f, h, g, G, g_1, g_2, ...$ are suitable scaling functions.

Thus below $p_c$ {\it for long times} we have
$$<r(t)> \; = \sum_s R_s s n_s g_1[(p_c-p)s^\sigma]$$
$$<r^2(t)> \; = \sum_s R^2_s s n_s g_2[(p_c-p)s^\sigma]$$
in $d$ dimensions. 

For $d>6$ the critical exponents should be those of the
Bethe lattice, with $\tau = 5/2, \; \sigma = \nu = 1/\delta = 1/2, \;
n_s \propto s^{-5/2} \exp[-{\rm const} (p_c-p)^2s]$. Also, $d_w=6,~x=3/2$.
Then

$$ <r(t)> \; \rightarrow {\rm Const} + \dots (p_c - p)^{1/2}$$
$$ <r^2(t)> \; \propto \log(p_c-p) + \dots$$
for $d > 6, \; p < p_c$. Right at $p=p_c$ the powers of $p_c-p$ are replaced by
the proper powers of $t \sim 1/(p_c-p)^3$:

$$ <r(t)> \; \rightarrow {\rm Const} + O(1/t^{1/6}) \eqno (1) $$
$$ <r^2(t)> \; \propto \log t$$
More precisely, in $ <r^2(t)> \; \propto \sum_s s^{-1} g(t/s^x, {\rm const})$ we
can approximate the sum by $\sum_s s^{-1}$ with an upper limit for $s$ of order
$t^{1/x}$, giving ln(const $t$) or Const + ln $t$.

For the third-leading term in $<r^2(t)>$ at $p=p_c$, we note that the leading
correction to scaling for $d>6$ comes from the leading irrelevant parameter
$w$, which represents the probability of having vertices with three 
bonds \cite{AGK}. This implies corrections to any leading power law behavior of
the form $(1+{\rm const}\times w^2X^{6-d})$, where $X$ represents an appropriate
length scale, which could be $r(t),~R_s$ or $t^{1/d_w}$. Specifically, at $p_c$
we expect corrections to $<r^2(t)>$ of relative order $t^{(6-d)/6}$, which become $1/t^{1/6}$ at $d=7$,  just
as for $<r(t)>$ above. The two corrections differ for $d>7$.
Thus finally at $d=7$  we expect:

$$  <r^2(t)> \; \propto \log t + {\rm Const} + O(1/t^{1/6}) \eqno(2)$$

\bigskip
Obviously, Eq. (1) is easier to test and reasonably confirmed by Fig. 2, using
$p_c = 0.088951$ from \cite{grass}. For Eq.
(2) we see in Fig. 3 a logarithmic long-time behaviour for $L=23$
(traditional method) and $L = 420$ (new method), while for short times and/or
small lattices deviations exist. Figure 4, similar to Fig. 2, shows that 
$<r^2(t)> \; = 1.25\ln(t)-7.8+8/t^{1/6}$
is consistent with our data for large systems over 8 decades in time. 
($<r^2(t=1) = p$ exactly, but the curve shown in Fig.4 extrapolates to a value
% remark about short times added
at $t=1$ differing from this value by 0.2,) The fit is
not shown in Fig. 3 since it would barely be distinguishable there from the 
curve for $L = 420$.  Less extensive simulations in eight dimensions, $L=180$,
nicely confirm $<r^2> \; \propto \log(t)$ + const, but are not accurate enough
to distinguish between correction exponents 1/3 and 1/6. In contrast, for seven
dimensions 1/6 fits over a wider range than 1/3. 
 
In summary, the new method gives results consistent with the old one but allows
for enormously larger lattice sizes; and these results are consistent with
scaling theory.

We thank Humboldt Foundation and GIF for supporting this collaboration, and 
R.M. Ziff, P. L'Ecuyer and P.M.C. de Oliveira for encouraging discussion.

\begin{figure}[hbt]
\begin{center}
\includegraphics[angle=-90,scale=0.6]{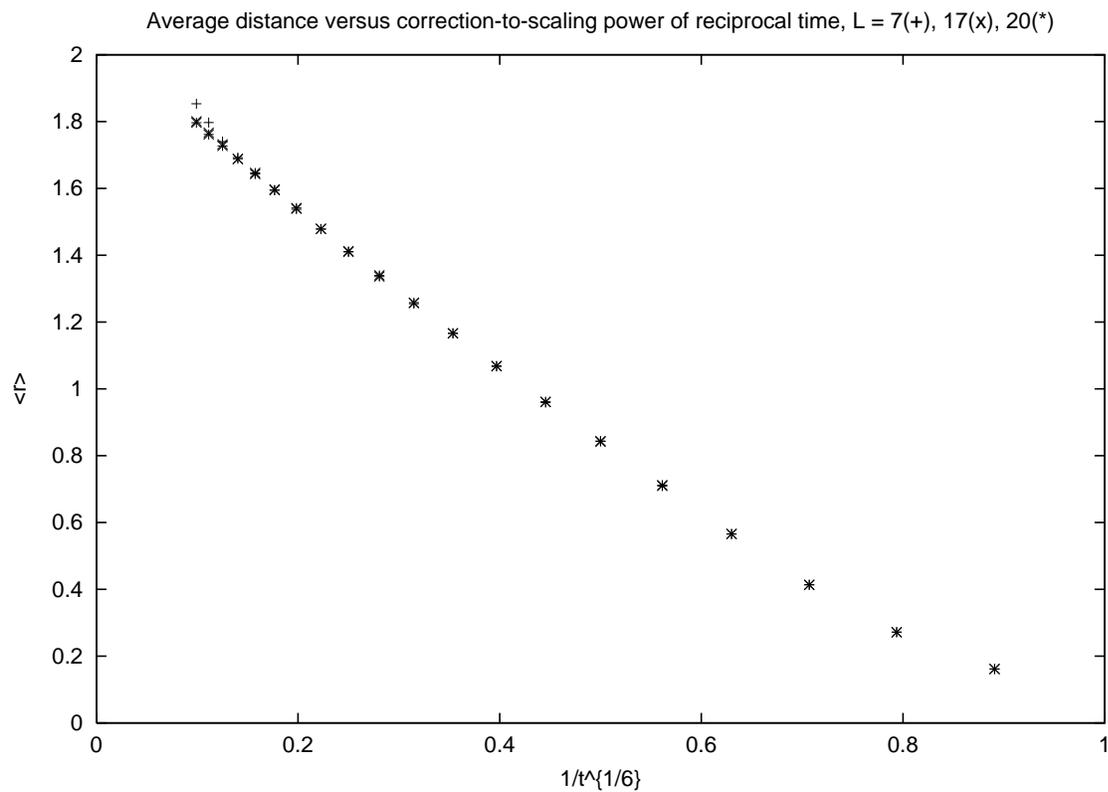}
\end{center}
\caption{ Average distance $<r>$ versus 1/time$^{1/6}$; asymptotically a 
straight line is expected leading to a finite intercept. }
\end{figure}

\begin{figure}[hbt]
\begin{center}
\includegraphics[angle=-90,scale=0.40]{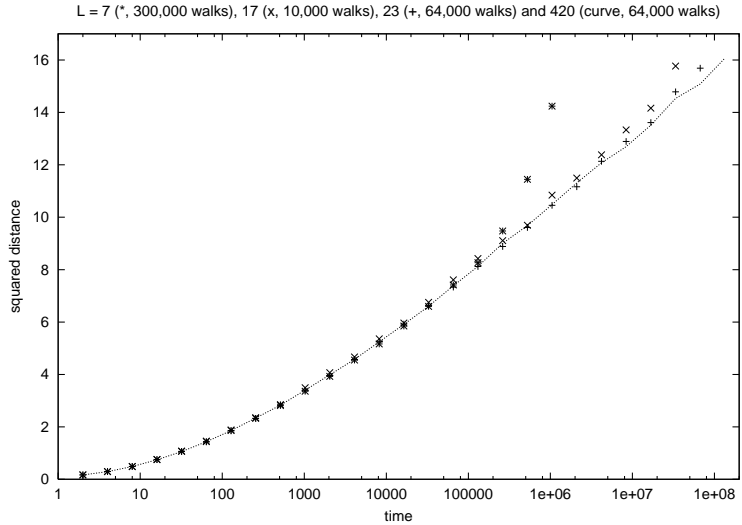}
\end{center}
\caption{ $<r^2(t)>$ versus time on logarithmic time scale; $L = 7$ and 17 
deviate for long times from $L = 23$ (traditional method) whereas $L = 420$
(continuous curve, new method) agrees with $L = 23$ for the observed times. }
\end{figure}

\begin{figure}[hbt]
\begin{center}
\includegraphics[angle=-90,scale=0.40]{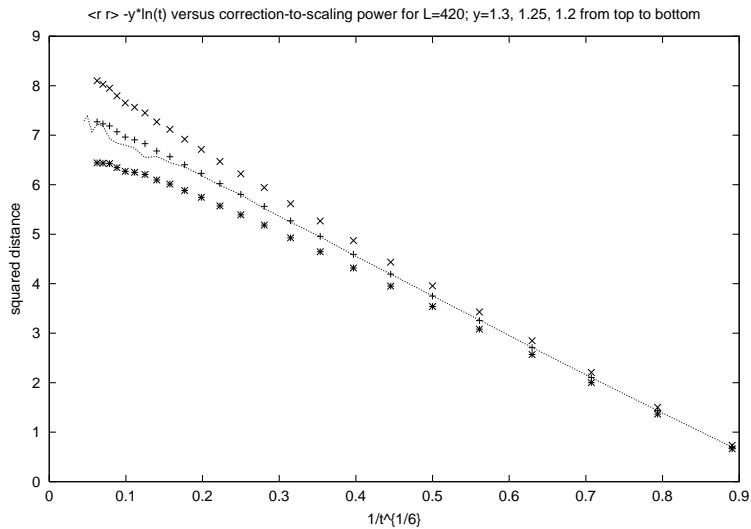}
\end{center}
\caption{$<r^2(t)> - y \ln(t)$ versus 1/time$^{1/6}$ for y = 1.2, 1.25 and 1.3;
(640,000 walks); the line comes from 64,000 walks to eight times longer $t$
up to 128 Megasteps. Asymptotically a 
% und corrected to up
straight line is expected leading to a finite intercept. }
\end{figure}

\end{document}